\documentclass[prl,twocolumn]{revtex4}

\begin{document}

\title{Temperature and Disorder Chaos in Low Dimensional Directed Paths}
\author{Rava A. da Silveira}
\affiliation{Lyman Laboratory of Physics, Harvard University, Cambridge, Massachusetts
02142, U.S.A. and\\
Laboratoire de Physique Th\'{e}orique, Ecole Normale Sup\'{e}rieure, 24 rue
Lhomond, 75005 Paris, France}
\author{Jean-Philippe Bouchaud}
\affiliation{Service de Physique de l'Etat Condens\'{e}, CEA--Saclay, 91191
Gif-sur-Yvette, France}

\begin{abstract}
The responses of a $1+\varepsilon $ dimensional directed path to temperature
and to potential variations are calculated exactly, and are governed by the
same scaling form. The short scale decorrelation (strong correlation regime)
leads to the overlap length predicted by heuristic approaches; its
temperature dependence and large absolute value agree with scaling and
numerical observations. Beyond the overlap length (weak correlation regime),
the correlation decays algebraically. A clear physical mechanism explains
the behavior in each case: the initial decorrelation is due to `fragile
droplets,' which contribute to the entropy fluctuations as $\sqrt{T}$, while
the residual correlation results from accidental intersections of otherwise
uncorrelated configurations.
\end{abstract}
\maketitle

Many disordered systems are said to be `complex' because they operate in the
vicinity of a large number of microscopically distinct, metastable states.
This generic property has several non-trivial manifestations, eminently,
slow dynamics and aging, irreversibility, and the relevance of preparation
history. A related aspect is the \textit{fragility}, or \textit{extreme
sensitivity}, of disordered systems: the `dominant' metastable states shift
from given microscopic configurations to completely different ones upon even
a slight variation of (external) parameters, a behavior that has been termed
`chaotic' as small perturbations map into major reconfigurations \cite%
{McKay,Fisher-Huse1986,Bray-Moore,Fisher-Huse1991}.

In experiments, temperature plays the role of a prototypical control parameter
as it is far easier to vary than, say, the microscopic details of the
disorder. \textit{Temperature chaos} (TC) stipulates that, rather
remarkably, the configurations probed by a disordered system (in its glassy
phase) at temperature $T$ are largely \textit{uncorrelated} to the ones
probed at a slightly different temperature $T+\delta T$, provided the system
is large enough, larger in fact than a characteristic \textit{overlap length}
$L^{\ast }$ (which for consistency must diverge as $\delta T\rightarrow 0$) 
\cite{Fisher-Huse1986,Bray-Moore,Fisher-Huse1991}. This scenario may provide
a natural explanation for several of the peculiar dynamical properties of
spin glasses, such as the absence of cooling rate effects and the
`rejuvenation' phenomenon \cite{Jonsson,Miyashita}. Indeed, if a
sufficiently large temperature step modifies drastically the equilibrium
configurations, it is equivalent in its effect to a quench from high
temperatures, regardless of the dwelling time at the high temperatures.
Numerical observations of dynamical rejuvenation in the absence of any
direct evidence of static TC \cite{Billoire,Berthier-Bouchaud}, however,
seem to point to alternative pictures which do not rely on an overlap length
but rather on the gradual freezing at successive scales \cite%
{Bouchaud2001,Berthier-Holdsworth,Berthier-Bouchaud}. While these are
compatible with the large value (in absolute terms) of $L^{\ast }$ inferred
from numerics \cite{Fisher-Huse1991}, as well as from mean-field \cite{Rizzo}
and Migdal-Kadanoff \cite{Aspelmeier} approximations, which would render the
overlap length irrelevant in the face of the small dynamical length scales in spin glasses
(a few tens of lattice spacings even on experimental time scales), recent
arguments suggest that TC may manifest itself in ways relevant to
experiments even at scales much smaller than $L^{\ast }$ \cite%
{Jonsson,Sasaki-Martin2003}.
From a theoretical point of view, systematic investigations of TC are
scarce. In the context of spin glasses, calculations have been confined to
mean-field \cite{Rizzo} and Migdal-Kadanoff \cite{Banavar,Aspelmeier}
approximations, begging for studies of simpler models that still contain the
physics of TC. Unfortunately, the exactly soluble Random Energy Model does
not display TC (except in the vicinity of the critical temperature \cite%
{Sales-Bouchaud}), unless it is extended to include explicit random entropy
variables \cite{Krzakala-Martin}. The next best candidate for a theoretical
treatment are directed paths (polymers) in a random potential. The
sensitivity of an array of pinned flux lines to potential and temperature
variations was studied by renormalization group methods \cite{Hwa-Fisher},
which yielded scaling forms (identical in both cases) in the strong
correlation regime. However, neither the dependence of the overlap length on
temperature nor the weak correlation regime were elucidated. In the case of
a single directed path, the (somewhat non-intuitive) result that changing
temperature is `equivalent' to changing disorder is confirmed by extensive
numerics (in Ref. \cite{Sales-Yoshino}, which also advances arguments based
on the mapping to bosons \cite{Kardar}). The present work characterizes TC
and disorder chaos (DC) for a directed path \textit{exactly}, by working
close to $d=1$ dimension, where a systematic $\varepsilon =d-1$ expansion
can be performed \cite{Derrida-Griffiths,Cook-Derrida}. To our knowledge,
this represents the first systematic, complete derivation of these effects;
our exact results broadly confirm and complement the predictions of scaling
arguments at strong correlation, elucidate the behavior at weak correlation,
and match a clear picture of the physical mechanisms involved in each regime.
Let us first summarize, in the context of directed paths, the scaling
arguments for TC and DC. The crucial assumption is that the free energy of a
directed path in a random potential consists, in the low temperature phase,
of the sum of a non-random term growing linearly with the length $L$ of the
path, with a prefactor regular in $T$, and, almost surely, a random
contribution $\Delta F$ of order $L^{\theta }$ with $\theta <1/2$. (For
example, $\theta =1/3$ holds exactly for two dimensional paths.) Now,
suppose that the random potential is slightly modified by a small unbiased
random addition of order $\delta V$ per bond, and assume $T=0$ for the sake
of simplicity. The resulting extra contribution to the energy of the
unperturbed ground state is obviously a random variable of order $\delta
VL^{1/2}$, which, for large $L$, exceeds the postulated $L^{\theta }$
fluctuations. To avoid a contradiction, a new ground state configuration,
roughly uncorrelated to the unperturbed one, must appear when $\delta
VL^{1/2}>L^{\theta }$, \textit{i.e.}, when $L>L^{\ast }\sim (\delta
V)^{-2/(1-2\theta )}$, the overlap length corresponding to \textit{potential
variations}. The case of \textit{temperature variations} is more subtle. At $
T>0$, configurations that do not optimize the energy contribute normal 
\textit{energy} fluctuations $\Delta E$, of order $L^{1/2}$, and similarly
for \textit{entropy} fluctuations $\Delta S$, such that the two $L^{1/2}$
contributions precisely cancel and give way to the expected $L^{\theta }$ 
\textit{free energy} fluctuations $\Delta F$. Now, if one could use the
thermodynamical identity $\partial \Delta F/\partial T=-\Delta S$, one would
conclude that the fluctuations of the free energy vary, upon a change of
temperature, by an amount $\Delta S\,\delta T\sim L^{1/2}\delta T$, which
again exceeds the `allowed' $L^{\theta }$ fluctuations whenever $L>L^{\ast
}\sim (\delta T)^{-2/(1-2\theta )}$. The only way out is that the `dominant'
configurations shift to radically different ones, inducing non-analyticities
in $\Delta F$ that can be seen as a sequence of micro-phase transitions at
all temperatures. That, in the glassy phase, disordered systems are in a
sense critical at all temperatures, and therefore extremely sensitive to
parameter changes, was suggested in the context of mean-field spin glasses 
\cite{Mezard-Parisi-Virasoro}. This generic fragility was predicted by
Fisher and Huse \cite{Fisher-Huse1986,Fisher-Huse1991}, as part of a rich
scaling picture from which the above arguments are extracted (see also Refs. 
\cite{Bray-Moore,Mezard,Hwa-Fisher}).
We now turn to our analytical results and their physical interpretation. (A
full account of the technical details will be published separately.) We
study the statistical mechanics of directed paths on a standard Berker lattice 
constructed recursively by replacing a bond by a diamond with $b$ two-bond branches. 
Each bond carries a Gaussian random energy with vanishing mean and
standard deviation $\sigma _{1}$ (though our results presumably hold more
generally, at least for sufficiently short tailed densities). To uncover the
effect of a small change in either temperature or random potential, we focus
on two paths at temperatures $T$ and $T^{\prime }$, and subjected to two
random potentials with a correlation coefficient $\rho_1$, and we
ask how they decorrelate as a function of their length $L$. Our 
calculation generalizes the beautiful work of Derrida and Griffiths \cite{Derrida-Griffiths} 
to non-vanishing temperatures \cite{Cook-Derrida} and to include more than one directed path.
We establish an exact recursion relation for the joint probability
distribution $\mathcal{P}\left( F(T),F^{\prime }(T^{\prime })\right) $ of
the free energies of the two directed paths. This recursion relation can be solved
in the low dimensional limit $b=1+\varepsilon \,$, in which $\mathcal{P}$ is close to a bivariate
Gaussian distribution.
The averages and higher (joint) moments of the free
energies may be extracted from this distribution; in line with our central
question, we focus on the normalized correlation%
\begin{equation}
\rho =\frac{\overline{\delta F(T)\,\delta F^{\prime }(T^{\prime })}}{\sqrt{%
\overline{\delta F(T)^{2}}\,\overline{\delta F^{\prime }(T^{\prime })^{2}}}},
\label{rho-def}
\end{equation}%
with $\delta F^{(\prime )}(T^{(\prime )})=F^{(\prime )}(T^{(\prime )})-%
\overline{F^{(\prime )}(T^{(\prime )})}$. The quantity $\rho $ codifies the
sensitivity of a directed path to temperature or random potential variations
The free energies of a pair of shortest possible paths are the bond energies
themselves, and thus $\rho (L=1)\equiv \rho _{1}$; $\rho _{1}=1$ 
if the two random potentials felt by the
directed paths are identical, while $0<\rho _{1}<1$ if
they are different (but correlated). For longer and longer paths, $\rho $
falls to zero due to thermal fluctuations (if $T\neq T^{\prime }$) or to
quenched fluctuations (if the random potentials are different). Our
calculation recovers the scaling form of the initial decorrelation in the
\textit{strong correlation regime}, and yields the overlap length $L^{\ast }$
(appearing here as the scale at which $\rho $ has fallen by a fraction, say $%
1/2$, of its initial value) in terms of the perturbation. In the \textit{%
weak correlation regime}, beyond $L^{\ast }$, we find that the residual
correlation $\rho $ decays much more slowly than it was previously
anticipated.
At short scales or strong correlation, the initial decorrelation evolves, as
a function $L$, according to%
\begin{equation}
\rho =1-\Delta L^{1-2\theta }+\mathcal{O}\left( L^{-\frac{1}{2}-2\theta
}\right) ,  \label{rho-initial}
\end{equation}%
where, in the limit $1 - \rho_1 \ll 1$ and $|\delta T| \ll \sigma_1$, %
\begin{equation}
\Delta =\left( 1-\rho _{1}\right) +\mathcal{A}\,s(T)\cdot \left( \frac{%
\delta T}{\sigma _{1}}\right) ^{2}.  \label{delta}
\end{equation}%
In these expressions,%
\begin{equation}
\theta =\frac{1}{2}-\frac{K_{2}}{\ln (2)}\varepsilon  \label{theta}
\end{equation}%
is the exponent ruling free energy fluctuations \cite%
{Derrida-Griffiths,Cook-Derrida}, 
\begin{equation}
s(T)=\frac{\pi ^{2}}{6\left( 2\sqrt{2}-1\right) }K_{1}\frac{T}{\sigma _{1}}
\label{entropy}
\end{equation}%
is the low temperature entropy per unit length, and the numerical constants $%
\mathcal{A}$, $K_{1}$, and $K_{2}$ take the values%
\begin{eqnarray}
\mathcal{A} &=&\frac{3}{\pi ^{2}}\int_{0}^{1}du\left\{ \frac{u\ln ^{2}\left(
u\right) }{\left( u+1\right) ^{2}}+2\frac{\ln ^{2}(u+1)}{u}\right\} \approx
0.23,  \nonumber  \label{constA} \\
K_{n} &=&-\int_{-\infty }^{\infty }du\,u^{n-1}E(u)\ln (E(u)).  \label{constK}
\end{eqnarray}%
($E(u)=\int_{-\infty }^{-u}\frac{dx}{\sqrt{2\pi }}e^{-x^{2}/2}$ denotes the
usual error function.) While our formulation is valid at any temperature,
the above explicit expressions correspond to the dominant low temperature
behavior. More generally, we find that $\Delta $ is a regular function of $%
(\delta T)^{2}$, at odds with `weak chaos' arguments \cite%
{Sales-Yoshino,Jonsson} which naively suggest an additional, singular $%
\left\vert \delta T\right\vert ^{3}$ contribution.

The above result for the initial decorrelation of a directed path in a
random potential may be viewed as follows. From the calculation, it appears
clearly that $\Delta $ embodies the `ignition' of the decorrelation, which
occurs dominantly at short scales. A path of length $L$ contains a number $%
\propto L$ of fragile `ignition droplets' responsible for the decorrelation
at short scales; their contribution to the overall decorrelation is
therefore proportional to $L$ divided by the square of the free energy
fluctuations, \textit{i.e.}, to $L^{1-2\theta }$, as in Eq. (\ref%
{rho-initial}) and in agreement with recent numerical data \cite%
{Sales-Yoshino} on directed paths in $d=1+1$ dimensions. Hence, $\Delta $ can be thought
of as the density of ignition droplets. (Equivalently, $\Delta $ corresponds
to the probability that an arbitrary small droplet takes part in the
`ignition.' Pushing this line of reasoning further, one can infer from the
calculation that a droplet of size $\ell $ takes part in the ignition with
probability $\propto 1/\ell ^{\theta }$, consistent with the dominance of
small droplets and the predictions of scaling theory \cite{Fisher-Huse1991}%
.) This interpretation becomes transparent in the case of TC if we consider,
for the sake of simplicity, $T^{\prime }=0$ and $T=\delta T$ small. Then the
correlation reads %
\begin{equation}
\rho \approx 1-\frac{1}{2}\left\{ \frac{T^{2}\overline{\left( S-\overline{S}%
\right) ^{2}}}{\overline{\left( E_{0}-\overline{E_{0}}\right) ^{2}}}-\left[ 
\frac{\overline{\left( E_{0}-\overline{E_{0}}\right) T\left( S-\overline{S}%
\right) }}{\overline{\left( E_{0}-\overline{E_{0}}\right) ^{2}}}\right]
^{2}\right\} ,  \label{rho-initial-approx}
\end{equation}%
where $E_{0}$ is the ground state energy of the path and $S=s(T)L$ its
entropy. Now, the entropy of the path at temperature $T$ comes from a number
of droplets, distinguishing nearly degenerate paths, such that the energy
difference between the arms of the droplet differ by less than $T$. These
droplets have a certain density per unit length, proportional to $s(T)$
(roughly, $s(T)/\ln (2)$). If we compare two samples, corresponding to two
different realizations of the random potential, some of the droplets
existing (at given positions on the path) in the first sample will not be
present in the second and \textit{vice versa}, so that the \textit{variance}
of the entropy too grows linearly with the length of the path. This picture
of Poissonian droplets implies furthermore that the \textit{variance} of the
entropy is given by the \textit{entropy itself}, \textit{i.e.}, we expect $%
\overline{\left( S-\overline{S}\right) ^{2}}\propto s(T)L\text{.}$ With $%
\overline{\left( E_{0}-\overline{E_{0}}\right) \left( S-\overline{S}\right) }%
\ll \overline{\left( S-\overline{S}\right) ^{2}}$ for large $L$ and the
scaling of ground state energy fluctuations $\overline{\left( E_{0}-%
\overline{E_{0}}\right) ^{2}}\sim \sigma _{1}^{2}L^{2\theta }$, Eq. (\ref%
{rho-initial-approx}) then reduces to%
\begin{equation}
\rho \approx 1-\frac{1}{2}s(T)\cdot \left( \frac{\delta T}{\sigma _{1}}%
\right) ^{2}\cdot L^{1-2\theta },  \label{rho-initial-approx-result}
\end{equation}%
in agreement with Eqs. (\ref{rho-initial},\ref{delta}). A similar argument
may be constructed in the case of DC.

A notable outcome of this discussion, substantiated by the analytic result,
is the fact that sample to sample variations of the entropy are of order $%
\sqrt{T}$ -- a behavior that extends to spin glasses on Berker lattices 
\cite{Aspelmeier}. As mentioned, this is due to the \textit{intermittent}
nature of the droplets contributing to entropy: a droplet is active with
probability $\propto T$, and inactive otherwise. Therefore all moments of
the entropy, including the mean and the variance, are proportional to $T$.

To conclude the discussion of the strong correlation regime, we note that,
according to Eq. (\ref{rho-initial}), once $L^{1-2\theta }$ becomes of order 
$1/\Delta $ the correlation has dropped by a significant fraction of its
initial ($L=1$) value. For TC, this observation defines an overlap length%
\begin{equation}
L^{\ast }\approx \left( \sqrt{\mathcal{A}\,s(T)}\frac{\delta T}{\sigma _{1}}%
\right) ^{-\frac{1}{1/2-\theta }}\sim \left( \sqrt{T}\delta T\right) ^{-%
\frac{1}{1/2-\theta }},  \label{overlap-length}
\end{equation}%
in agreement with the outcome of Fisher and Huse's scaling theory \cite%
{Fisher-Huse1991}. Inserting numerical values, we find a rather large
overlap length, $L^{\ast }\approx 10^{4}$, even for $T=\sigma _{1}$, $\delta
T=\sigma _{1}/2$, and $\theta =1/3$ (the exact value in $d=1+1$ dimensions), in
qualitative agreement with numerical observations for both directed paths 
\cite{Fisher-Huse1991,Sales-Yoshino} and spin-glasses \cite{Aspelmeier}.
Finally, since the (small) fragile droplets dominate the ignition mechanism
for TC, consistency requires that $L^{\ast }$ be much larger than the
typical distance $\ell _{T}\propto 1/T$ between droplets. From Eq. (\ref%
{overlap-length}), this is indeed the case as long as $0<\theta <1/2$.

In the weak correlation regime with $L\gg L^{\ast }$, we find, at odds with
common expectations, a slow algebraic decay%
\begin{equation}
\rho \sim L^{-\frac{1-2K_{2}}{\ln (2)}\varepsilon }  \label{rho-residual}
\end{equation}%
for both temperature and disorder decorrelation. Again, a simple physical
picture explains this behavior. As confirmed by our calculation, at large $L$
the ground state dominates the statistics or, equivalently, the temperature
effectively scales to zero. Thus, whether at slightly different temperatures
or in slightly different random potentials, our two directed paths behave at
large scales as if \textit{pinned} by two \textit{uncorrelated} random energy
configurations, and their residual correlation arises from the (rare) bonds
at which they intersect. Whence we expect%
\begin{equation}
\rho =\frac{\overline{\left( E_{0}-\overline{E_{0}}\right) \left(
E_{0}^{\prime }-\overline{E_{0}}\right) }}{\overline{\left( E_{0}-\overline{%
E_{0}}\right) ^{2}}}\approx \frac{\mathcal{I}(L)\overline{\left( e_{\mathcal{%
I}}-\overline{e_{\mathcal{I}}}\right) ^{2}}}{\overline{\left( E_{0}-%
\overline{E_{0}}\right) ^{2}}},
\end{equation}%
where $\mathcal{I}(L)$ is the average number of intersection bonds and $e_{%
\mathcal{I}}$ the energy of an intersection bond. Now, for completely
uncorrelated paths on a Berker lattice with branching $b$, $\mathcal{I}%
(L)=L^{1-\frac{\ln (b)}{\ln (2)}}$ trivially. Extrapolating this result to $%
b=1+\varepsilon $, we note that the quantity 
\begin{equation}
\frac{\mathcal{I}(L)}{\overline{\left( E_{0}-\overline{E_{0}}\right)
^{2}}}\sim \frac{L^{1-\frac{\ln (b)}{\ln (2)}}}{L^{2\theta }}\sim L^{-\frac{
1-2K_{2}}{\ln (2)}\varepsilon }
\end{equation}%
precisely coincides with the analytic result of Eq. (\ref{rho-residual}).
From this argument, we conclude that the intersections occur indeed
`accidentaly': free energy minimization manifests itself, at best,
subdominantly.
The physical mechanism for residual correlation just described does not
bear on any specificity of the model and affords us with natural conjectures
for the large $L$ decorrelation of directed paths in higher dimensions. For
two paths in $d=1+D$ dimensions, emerging from the same end point but otherwise
uncorrelated (on scales $L\gg L^{\ast }$), the probability of encounter at a
distance $\ell $ away from the end point is $1/\ell ^{\zeta D}$, where $%
\zeta =\left( 1+\theta \right) /2$ is the wandering exponent. Therefore, the
mean number of accidental intersections, at large $L$, grows like $%
L^{1-\zeta D}$ if $\zeta D<1$ and converges to a constant if $\zeta D>1$, so
that we expect the correlation to scale as $\rho \sim L^{1-\zeta D-2\theta }$
if $\zeta D<1$ and $L^{-2\theta }$ if $\zeta D>1$. For example, in $
d=1+1$ dimensions, $\zeta =2/3$ and $\theta =1/3$, whence $\rho \sim L^{-1/3}$ for $%
L\gg L^{\ast }$ (instead of the previously conjectured $L^{-4/3}$ decay \cite%
{Fisher-Huse1991}). It is unfortunately impossible to check this prediction
against recent numerical results \cite{Sales-Yoshino}, as these do not
extend to the asymptotic decorrelation regime.

In sum, while the decorrelation mechanisms -- entropic fluctuations of
fragile droplets for TC and energetic competition for DC -- are different,
the scaling form of the free energy correlation, when
expressed in terms of $L/L^{\ast }$, is exactly the same in the two 
cases. We calculated exactly
this scaling form and the dependence of the overlap length $L^{\ast }$ on
either temperature or potential variations. Perhaps surprisingly, the free
energy correlation decays as a weak power law, rather than exponentially, at
large $L$. A study of the statistics of the overlap between unperturbed and
perturbed paths, possibly using similar methods, would be an interesting
extension of the present work to better understand the crossover from short
to long scales and the nature of the overlap length. In another worthwhile
extension, one could apply the systematic approach on low dimensional Berker
lattices to study TC in spin glasses. However, beyond its added
calculational complication, the case of spin glasses might be more subtle
due to the presence of several exponents ruling the glassy phase \cite{BKM},
instead of a single one, $\theta $, for directed paths. Physically, the
presence of several exponents reflects the fact that the scaling of energy
fluctuations depends on the topological nature of excitations (that may be
compact, sponge-like, or fractal), and it might well be that on Euclidean
lattices a backbone of `strong links' prevents a complete decorrelation for
small temperature changes. 
An interesting variation on the directed path problem, that may mimic this
mechanism, incorporates a fat tailed random potential $V$ (distributed
according to a power law $P(V)$), as in that case the directed path
configurations are controlled by the particularly favorable sites. In $d=1+1$
dimensions, a simple argument suggests that TC exists only if $P(V)$ decays
slower than $V^{-9/2}$. From a theoretical point of view, a finer
understanding of such a behaviour is, doubtless, worthwhile.

We thank E. Bertin, G. Biroli, B. Derrida, T. Emig, D.S. Fisher, O. Martin,
and H. Yoshino for useful discussions. R.A.S. is grateful to D. Danielou for
his hospitality at the CDF in Paris, where this work was pursued in part,
and to A.N. Berker for his lattices. This work was supported by a \textit{%
Chateaubriand} fellowship, the Harvard Society of Fellows, and a Young
Researcher Fellowship from the \textit{Fonds national suisse} (R.A.S.).

\end{document}